\documentclass[aps,prl,twocolumn,superscriptaddress]{revtex4}
\usepackage{amsmath}
\usepackage{graphicx}
\usepackage{graphics}
 
\begin{document}

\newcommand{\diff}[2]{\frac{\partial #1}{\partial #2}}
\newcommand{\der}[2]{\frac{d #1}{d #2}}
\newcommand{\diffr}[1]{\diff{#1}{r}}
\newcommand{\diffth}[1]{\diff{#1}{\theta}}
\newcommand{\diffz}[1]{\diff{#1}{z}}
\newcommand{\secdiff}[2]{\frac{\partial^2 #1}{\partial #2^2}}
\newcommand{\twovec}[2]{\left(\begin{array}{c} 
#1 \\ #2 \end{array}\right)}
\newcommand{\threevec}[3]{\left(\begin{array}{c} 
#1 \\ #2 \\ #3 \end{array}\right)}
\newcommand{\twomatrix}[4]{\left(\begin{array}{cc} 
#1 & #2 \\ #3 & #4 \end{array}\right)}
\newcommand{\threematrix}[9]{\left(\begin{array}{ccc} 
#1 & #2 & #3 \\ #4 & #5 & #6 \\ #7 & #8 & #9 \end{array}\right)}
\newcommand{\twomatrixdet}[4]{\left|\begin{array}{cc} 
#1 & #2 \\ #3 & #4 \end{array}\right|}
\title{A general theoretical description of N-body recombination}
\author{N. P.~Mehta}
\affiliation{Department of Physics and JILA, University of Colorado, Boulder, CO 80309, USA}
\affiliation{Grinnell College, Department of Physics, Grinnell, IA 50112}
\email{mehtan@jilau1.colorado.edu}
\author{Seth T. Rittenhouse}
\affiliation{Department of Physics and JILA, University of Colorado, Boulder, CO 80309, USA}
\email{rittenhouse@colorado.edu}
\author{J.~P.~D'Incao}
\affiliation{Department of Physics and JILA, University of Colorado, Boulder, CO 80309, USA}
\affiliation{Institut f\"ur Quantenoptik und Quanteninformation, \"Osterreichische Akademie der Wissenschaften, 6020 Innsbruck, Austria}
\email{jpdincao@jila.colorado.edu}
\author{J.~von Stecher}
\affiliation{Department of Physics and JILA, University of Colorado, Boulder, CO 80309, USA}
\email{Javier.Vonstecher@colorado.edu}
\author{Chris~H.~Greene}
\affiliation{Department of Physics and JILA, University of Colorado, Boulder, CO 80309, USA}
\email{chris.greene@colorado.edu}
\date{\today}

\begin{abstract}
  A formula for the cross section and event rate constant describing
  recombination of $N$ particles are derived in terms of general
  $S$-matrix elements.  Our result immediately yields the generalized
  Wigner threshold scaling for the recombination of $N$ bosons.  A
  semi-analytic formula encapsulates the overall scaling with energy
  and scattering length, as well as resonant modifications by the
  presence of $N$-body states near the threshold collision
  energy in the entrance channel.  We then apply our model to the case
  of four-boson recombination into an Efimov trimer and a free atom.
\end{abstract}

\maketitle

Few-body processes have played an increasingly crucial role in the
physics of strongly interacting quantum gases.  Three-body
recombination in particular~\cite{ThreeBodRecombPapers} contributes
strongly to atom-loss and is primarily responsible for controlling the
lifetime of Bose-Einstein condensates (BEC) because the kinetic energy
released in the reaction is usually sufficient to eject the collision
partners from the trapping potential~\cite{BraatenHammer2006PhysRep}.
At ultracold temperatures, the physics of three-body recombination is
largely controlled by the effective long-range potential in the
entrance (three-body) channel.  Hence, a threshold analysis of these
potentials immediately yields information about the scaling
behavior of the event rate constant $K_3$ with respect to both the
energy $E$ and the (s-wave) scattering length $a$.  For instance, a
threshold analysis shows that $K_3$ scales as $a^4$ and $E^0$ for
bosons~\cite{ThreeBodRecombPapers}, making three-body recombination an
important process in the ultracold limit.  
In contrast, the long-lived
nature of the ultracold polarized Fermi gas is explained by the $E^2$
scaling of $K_3$ near threshold~\cite{SunoEtAlPRL2003}.  
Further, the
presence of Efimov states~\cite{Efimov} near the threshold collision
energy resonantly enhances recombination~\cite{ThreeBodRecombPapers}.  This resonant feature was
exploited to confirm the existence of Efimov states in experiments
with ultracold cesium~\cite{KraemerEtAl2006Nature}.  The increasing
number of experiments observing Efimov
physics~\cite{MoreRecentEfimovExperiments} highlights the importance
of few-body physics in our understanding of strongly interacting
quantum gases.

The natural extension of three-body recombination is to four-bodies.
Considerable progress has in fact been made towards the calculation of
four-body processes, notably collision cross sections involving
two-body fragmentation channels~\cite{FourBodScatTheory} and bound state energies~\cite{PlatterEtAl2004PRA}.  Scattering
processes involving four or more free atoms are far more complex and
require a deep analysis of the multiparticle
continua~\cite{VonStecherEtAl2008NatPhys, WangEsry2009PRL}.
Nevertheless, recent studies have proposed four-body recombination as
an efficient process for the production of Efimov
trimers~\cite{WangEsry2009PRL,VonStecherEtAl2008NatPhys}.  Four-boson
borromean states (four-body bound states with no bound subsystems)
with universal properties tied to Efimov physics were 
recently predicted~\cite{HammerPlatter2007EPL}, further explored~\cite{VonStecherEtAl2008NatPhys}, and measured by the
Innsbruck group~\cite{FerlainoEtAl2009PRL} in atom loss due to
recombination.  The present analysis is concerned with the 
recombination of $N$ particles into a channel with $N-1$ particles bound together and one free.  

We use hyperspherical coordinates
(see~\cite{SmirnovShitikova} and~\cite{Avery} for useful reviews), in
which Jacobi vectors $\vec{\rho}_i$ are transformed to a set of
angular coordinates collectively denoted $\Omega$, plus a radial
coordinate called the hyperradius $R$ defined by
$\mu_NR^2=\sum_{i=1}^{N-1}{\mu_i\rho_i^2}$, where $\mu_{N}= [(\prod_i
m_i)/M]^{1/(N-1)}$ is the $N$-body reduced mass, $M=\sum_i m_i$ is the
total mass of the system and $\mu_i$ is the reduced mass associated
with the $i^{\text{th}}$ Jacobi vector.  At large $R$, the solutions
to the angular portion of the Schr\"odinger equation yield the
fragmentation channels of the $N$-body system, and the quantum numbers
labeling those solutions index the $S$-matrix.

\emph{Derivation of the generalized cross section.}---This formulation begins by considering scattering by a purely
hyperradial potential in $d$-dimensions, and then obtains the
generalized cross section \textquotedblleft by
inspection\textquotedblright.  For clarity, we adopt a notation that
resembles the usual derivation in three dimensions.

In $d$-dimensions, the wavefunction at large $R$ behaves as:
\begin{equation}
  \Psi ^{I}\rightarrow e^{i\vec{k}\cdot \vec{R}}+f(\hat{k},\hat{k}^{\prime })%
  \frac{e^{ikR}}{R^{(d-1)/2}}
\end{equation}%
Equivalently, an expansion in hyperspherical harmonics is written in
terms of unknown coefficients $A_{\lambda \mu}$:
\begin{equation}
\label{psi2}
  \Psi ^{II}=\sum_{\lambda ,\mu }{A_{\lambda \mu }Y_{\lambda \mu }(\hat{R}%
    )[j_{\lambda }^{d}(kR)\cos \delta _{\lambda }-n_{\lambda }^{d}(kR)\sin
    \delta _{\lambda }]}
\end{equation}%
Here, $Y_{\lambda \mu}$ are hyperspherical harmonics (solutions to the free-space angular equation $(\Lambda^2 - \lambda(\lambda+d-2))Y_{\lambda \mu}=0$, where $\Lambda^2$ is the grand angular momentum operator~\cite{Avery}) and
$j_{\lambda}^d$ ($n_{\lambda}^d$) are hyperspherical Bessel (Neumann)
functions~\cite{Avery}.

Identification of the incoming wave parts of $\Psi ^{I}$ and $\Psi
^{II}$ yields the coefficients $A_{\lambda \mu }$, whose insertion into Eq.~(\ref{psi2})
gives:
\begin{align}
  f(\hat{k},\hat{k}') = \left(\frac{2 \pi}{k}\right)^{\frac{d-1}{2}}&\sum_{\lambda \mu}{}i^\lambda e^{-i(d/2-1+\lambda)\pi/2 - i\pi/4} \notag\\
  &\times Y_{\lambda
    \mu}^*(\hat{k})Y_{\lambda\mu}(\hat{k}')(e^{2i\delta_\lambda}-1).
\end{align}
The immediate generalization of this elastic scattering amplitude to
an anisotropic short-range potential is:
\begin{align}
  f(\hat{k},\hat{k}') = \left(\frac{2 \pi}{k}\right)^{\frac{d-1}{2}}&\sum_{\lambda \mu \lambda' \mu'}{}i^\lambda e^{-i\eta_\lambda } Y_{\lambda \mu}^*(\hat{k})Y_{\lambda^{\prime}\mu^{\prime}}(\hat{k}^{\prime}) \notag \\
  & \times
  (S_{\lambda\mu,\lambda^{\prime}\mu^{\prime}}-\delta_{\lambda\lambda^{\prime}}\delta_{\mu\mu^{\prime}}).
\end{align}
where $\eta_\lambda = (d/2-1+\lambda)\pi/2 +\pi/4.$ Upon integrating
$\vert f(\hat{k},\hat{k}^{\prime }) \vert^2$ over all final
hyperangles $\hat{k}$, and {\it averaging} over all initial
hyperangles $\hat{k}^{\prime}$ as would be appropriate to a gas phase
experiment, we obtain the average integrated elastic scattering cross
section by a short-range potential:
\begin{equation}
  \label{sigmadist}
  \sigma^{dist}=\left( \frac{2\pi }{k}\right) ^{d-1} %
  \frac{1}{\Omega(d)}
  \sum_{\lambda \mu \lambda^{\prime} \mu^{\prime}}
  { \left\vert S_{\lambda\mu,\lambda^{\prime}\mu^{\prime}}-\delta_{\lambda\lambda^{\prime}}\delta_{\mu\mu^{\prime}}%
    \right\vert {^{2}}}
\end{equation}%
where $\Omega(d)=2\pi^{d/2}/\Gamma(d/2)$ is the total solid angle in
$d$-dimensions~\cite{Avery}.  This last expression is immediately
interpreted as the generalized average cross section resulting from a
scattering event that takes an initial channel into a final channel,
$i \equiv \lambda^{\prime}\mu^{\prime} \rightarrow \lambda\mu \equiv f
$.  Since this S-matrix is manifestly unitary in this representation,
it immediately applies to inelastic collisions as well, including
$N$-body recombination.  It is worth noting that the sum in
Eq.~(\ref{sigmadist}) should include degeneracies, and that the cross
section should be appropriately averaged over initial spin substates
and collision energies.  

In this form, we can \emph{interpret} the generalized cross section derived above in terms of the unitary S-matrix computed by solving the exact coupled-channels reformulation of the few-body problem within the adiabatic hyperspherical representation~\cite{MacekJPB1968}.
In principle this can describe collisions of an arbitrary number of particles.  
Identical particle symmetry is handled by summing over all indistinguishable amplitudes before taking the square, averaging over incident directions and momenta for a given energy, followed by integrating over distinguishable final states to obtain the total cross section: $\sigma^{indist}=N_p \sigma^{dist}$.
Here $N_{p}$ is the number of terms in the permutation
symmetry projection operator (e.g. for $N$ identical particles, $N_{p}=N!$.) 

The cross section for total angular momentum $J$ and parity $\Pi$ includes an explicit $2J+1$ degeneracy.
Hence, the total generalized cross section (with dimensions of length$^{d-1}$) for $N$ particles in all incoming channels $i$ to scatter into the final state $f$, 
properly normalized for identical particle symmetry, is given in terms of general $S$-matrix elements as
\begin{align}
\label{sigmaindist}
\sigma^{indist}_{f i}({J^{\Pi}})&=N_{p}  \left(\frac{2\pi }{k_i}\right) ^{d-1} \frac{1}{\Omega(d)} \notag \\
\times &\sum_{i } 
\left( 2J+1\right) \left\vert {S^{J^{\Pi}}_{fi }-\delta _{fi}}%
\right\vert {^{2}}
\end{align}%
The event rate constant [recombination probability per second for each distinguishable $N$-group within 
a (unit volume)$^{(N-1)}$] is the generalized cross section Eq.~(\ref{sigmaindist}) multiplied by a factor of the 
$N$-body hyperradial \textquotedblleft velocity\textquotedblright\ (including factors of $\hbar$ to explicitly show the units of $K_N$):
\begin{equation}
K^{J^{\Pi}}_{N}=\frac{\hbar k_i}{\mu _{N}}\sigma^{indist}_{f i}({J^{\Pi}}).
\end{equation}%

\emph{Treatment of $N$ bosons.}--- 
The structure of a relevant $S$-matrix element from an adiabatic hyperspherical viewpoint is seen to be 
$S_{fi}^{J^{\Pi}} = \langle \Phi^{J^{\Pi}}_f(R;\Omega)F_f(R)|\hat{S}| \Phi^{J^{\Pi}}_i(R;\Omega)F_i(R)\rangle$, 
where $\Phi^{J^{\Pi}}_i$ and $\Phi^{J^{\Pi}}_f$ are the channel functions (i.e. solutions to the hyperangular part of the Schr\"odinger equation
in the limit $R\rightarrow \infty$) for the entrance and final channel respectively.  For all $N$-body entrance channels we have $i \rightarrow \lambda$, where $\lambda$ is the hyperangular momentum quantum number associated with eigenfunction 
$Y_{\lambda \mu}$.   The functions $F_i$ and $F_f$ are the large $R$ solutions to the
coupled hyperradial equations obtained in the adiabatic hyperspherical representation~\cite{MacekJPB1968} (in units where $\hbar=1$):
\begin{equation}
\left[\frac{-1}{2\mu_{N}}\secdiff{}{R} + W_i(R) - E\right]F_i + \sum_{f \ne i}V_{i f}(R)F_{f}=0,
\end{equation}  
Asymptotically (as $R\rightarrow\infty$) the couplings $V_{i f}(R)$ vanish and the effective potentials in the N-body continuum channels
are expressed in terms of an \emph{effective} angular momentum quantum number $l_e$:
\begin{equation}
\label{Wlong} 
W_\lambda(R)  \rightarrow \frac{l_e(l_e+1)}{2\mu_{N}R^2} \;\;\text{with} \;\; l_e=(2\lambda+d-3)/2.
\end{equation}
Near threshold, the recombination cross section is controlled by the \emph{lowest} $N$-body entrance channel 
$i=\lambda\rightarrow \lambda_{min}=0$. 
For $N$ identical bosons in a thermal (non-quantum-degenerate) gas cloud, the dominant contribution to Eq.~(\ref{sigmaindist}) 
is from $J^\Pi=0^+$.  Further, unitarity of the $S$-matrix allows us to write the \emph{total} rate constant 
summed over all final channels (using $d=3N-3$) as:
\begin{equation}
\label{KNthresh}
K^{0^{+}}_{N} =\frac{2 \pi \hbar N! }{\mu _{N}} 
\frac{(2\pi/k)^{(3N-5)}}{\Omega(3N-3)}
\left(1-\left\vert S^{0^+}_{00}\right\vert {^{2}}\right).
\end{equation}%

\begin{figure}[!t]
\begin{center}
\leavevmode
\includegraphics[width=2.8in, clip=true]{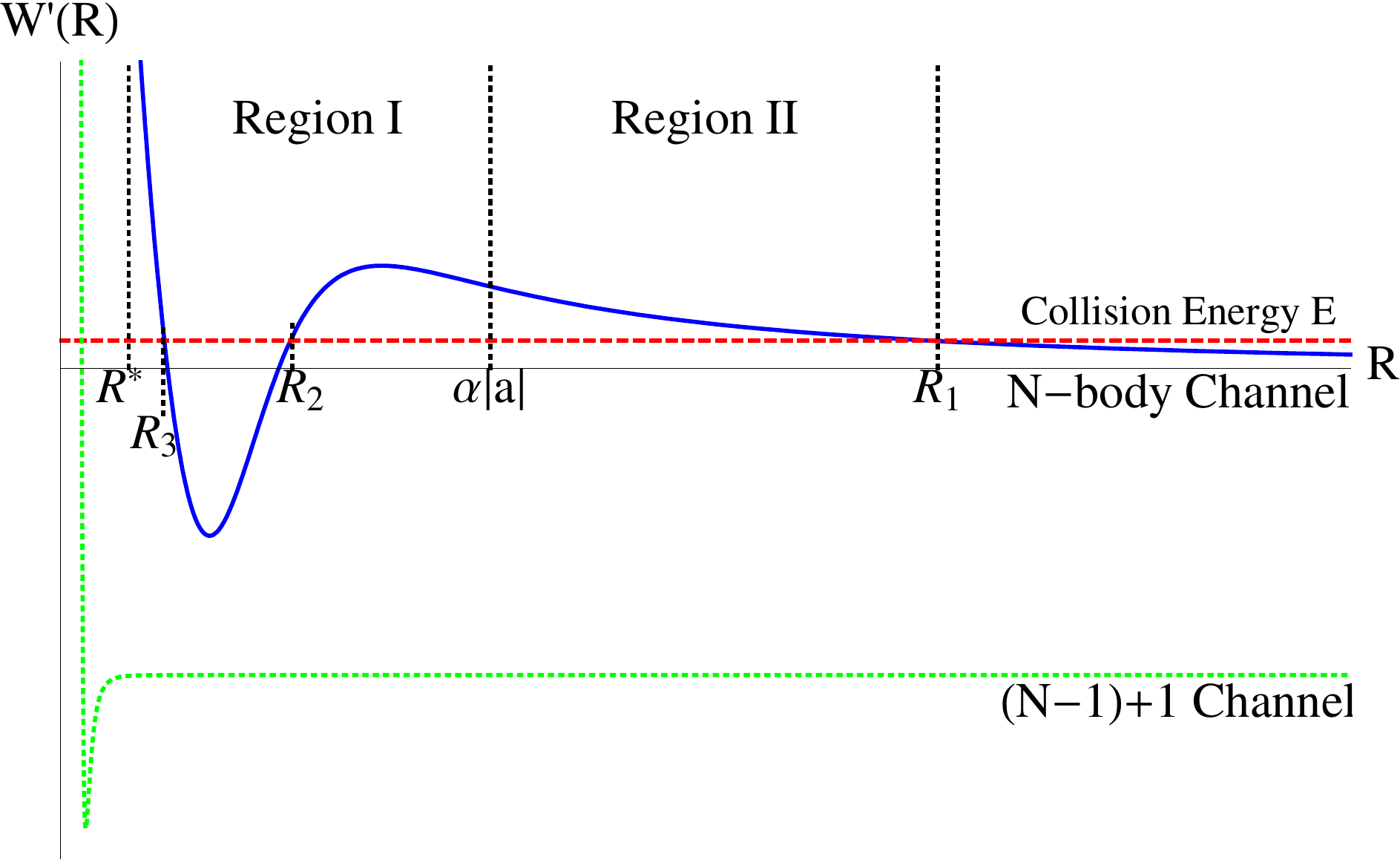}
 \begin{picture}(0,0)(0,0)
      \put(-132,90){$W'\rightarrow\frac{(l_e+1/2)^2}{2\mu_NR^2}$}
 \end{picture}
\caption{(color online) A schematic representation of the $N$-boson hyperradial potential curves is shown.   When a metastable $N$-boson state crosses the collision energy threshold at $E=0$, 
$N$-body recombination into a lower channel with $N-1$ atoms bound plus one free atom is resonantly enhanced.}
\label{fig:NBosonSchematic}
\end{center}
\end{figure}
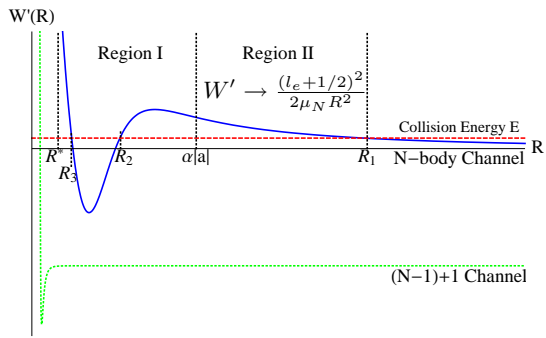

Based on prior hyperspherical treatments for three~\cite{DincaoEsry2005PRL} and four~\cite{VonStecherEtAl2008NatPhys} bosons, in cases where 
one or more $(N-1)$-body state is bound with no bound states of fewer particles,
we expect the hyperradial potentials to schematically look like those sketched in 
Fig.~\ref{fig:NBosonSchematic}.  Monte Carlo calculations~\cite{HannaBlume2006PRA} of $N$-boson systems provide 
evidence of super-borromean states which could produce the topology shown in Fig.~\ref{fig:NBosonSchematic}.
For four bosons, potentials with the topology shown occur at negative scattering lengths ($a<0$).

The semiclassical (WKBJ) treatment of Berry~\cite{BerryProcPhysSoc1966} can be generalized, giving a 
\emph{complex} phase shift $\delta_{l_e}$ for collisions of $N$ identical bosons.
The phase shift is acquired in terms of the phase $\phi$ accumulated in the classically allowed region 
($R_3<R<R_2$), and the tunneling integral $\gamma$ in the classically forbidden regions: 
\begin{equation}
\phi=\int_{R_3}^{R_2}{q(R)dR}; \;\; \text{and} \;\;
\gamma=\text{Im}{\int_{R^*}^{(3N-5)/2k}q(R)dR},
\end{equation}
where $q(R)=\sqrt{2\mu_{N}(E-W'(R))}$ ($W'$ is the effective potential including the Langer correction~\cite{MorseFeshbach}), 
and $R^*$ is the hyperradius at which the inelastic coupling to the exit channel peaks
(typically at hyperradii much greater than the range of the two-body interaction $r_0$; i.e. $r_0\ll R^*\lesssim R_3$ for recombination into a universal $(N-1)$-body state.)

The inelastic amplitude is incorporated into the phase by letting $\phi \rightarrow \phi +i\eta$~\cite{BraatenHammer2004PRA}.
Applying the connection formulas for each of the classical turning points $R_1$, $R_2$ and $R_3$ shown in Fig.~\ref{fig:NBosonSchematic} gives the phase shift:
\begin{equation}
\delta_{l_e}=\delta_{l_e}^{(0)}-\arctan{\left(\frac{1}{4}e^{-2\gamma}\cot{(\phi+i\eta-\pi/2)}\right)}
\end{equation}
Here, $\delta_{le}^{(0)}$ is the real semiclassical phase shift derived by ignoring the interior region, which does not contribute to the inelastic probability. 
The $S$-matrix element describing scattering from one $N$-boson state to another is then simply:
$S_{00}=e^{2i\delta_{l_e}}$, and the total probability to recombine into all available final channels is:
\begin{equation}
1-|S_{00}|^2 =\frac{e^{-2\gamma}}{2}\frac{\sinh{(2\eta)}}{\cos^2{\phi}+\sinh^2{\eta}}A(\eta,\gamma,\phi)
\end{equation}
where the function $A(\eta,\gamma,\phi)$:
\begin{equation}
%\label{}
A(\eta,\gamma,\phi)=\left|1+\frac{e^{-2\gamma}}{4}\tanh{(\eta+i\phi)}  \right|^{-2}
\end{equation}
is equal to unity in the threshold limit \emph{unless} both $\eta\rightarrow 0$ and $\phi\rightarrow \pi/2$.  

The threshold energy dependence of $K_N^{0^+}$ is found by breaking $\gamma$ into two pieces corresponding to the regions shown in Fig.~\ref{fig:NBosonSchematic}:
\begin{equation}
\gamma_{II}=\int_{\alpha |a|}^{(3N-5)/2k}{|q(R)|dR}; \;\;\;\; \gamma_{I}=\text{Im}{\int_{R^*}^{\alpha|a|}{q(R)dR}}.
\end{equation}
For recombination into a \emph{universal} $(N-1)$-body bound state, we expect $R^*\gg r_0$, in which case 
$\gamma$, $\phi$ and $\eta$ are also universal.
A simple calculation shows 
$e^{-2\gamma_{II}}=(2k \alpha  |a|/(3N-5))^{3N-5}$.
It is convenient to introduce an $\alpha$-independent function $C(a)=C_N\alpha^{(3N-5)}e^{-2\gamma_{I}}$.
The constant $C_N$ must be adjusted to give the correct overall scale of $K_N^{0^+}$.  
The full $N$-boson recombination rate constant in the ultracold limit is then:
\begin{equation}
\label{Kn}
K^{0^{+}}_{N} =\frac{\pi \hbar N!}{\mu _{N}\Omega(3N-3)} \left(\frac{4 \pi  |a|}{3N-5}\right)^{3N-5}
\frac{C(a) \sinh{(2\eta)} }{\cos^2{\phi}+\sinh^2{\eta}} 
\end{equation}
Note that this is a quantitative result valid in the threshold regime.  
It is a \emph{constant} (independent of $k$), and scales roughly as $|a|^{3N-5}$ (in agreement with~\cite{ThogersenEtAl2008EPL}).
Hence,  $N$-body processes could contribute to the total atom loss at higher particle density $n$ through terms of the form $-K_Nn^N$:
$\der{n}{t}=\sum_{N=2}^{N_{max}}{-L_N n^{N}}$,
where the atom loss rate $L_N$ is related to the event rate by $L_N = K_N/(N-1)!$ 
provided each recombination event ejects $N$ atoms from the trapped gas.  
(Recall also that for a quantum-degenerate Bose gas, the above expressions for $K_N$ must be divided by $N!$~\cite{kagan1985ebc}.)  
For $N=3$, Eq.~(\ref{Kn}) is 
consistent with the expression found by Braaten and Hammer~\cite{BraatenHammer2004PRA} 
and the phase $\phi$ acquires the universal log-periodicity 
characteristic of Efimov physics~\cite{Efimov}.

\begin{figure}[!t]
%\begin{picture}
\begin{center}
\leavevmode
\includegraphics[width=3.0in, clip=true]{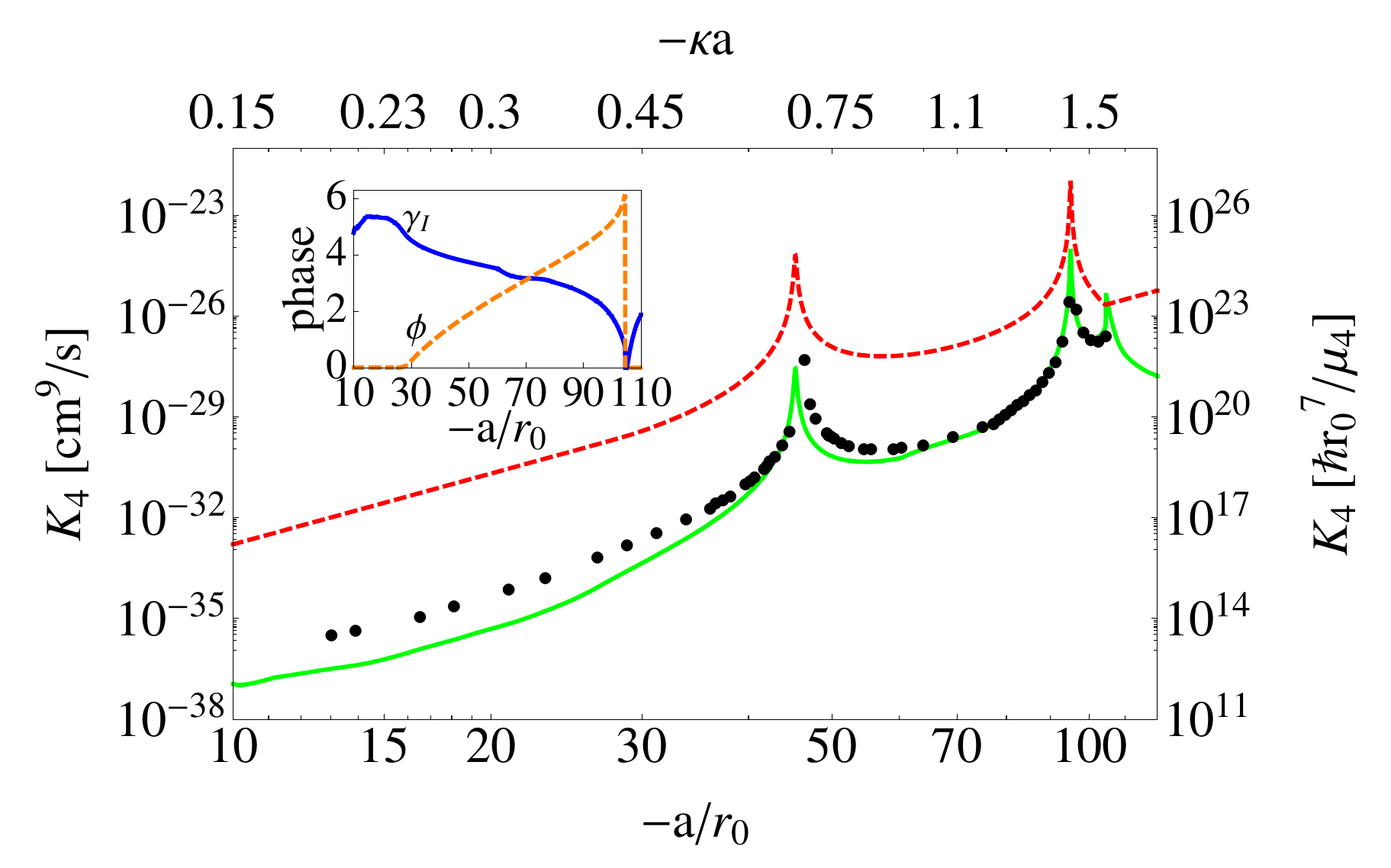}
\caption{(color online) The four-boson recombination rate constant $K_4^{0^{+}}$ is shown over approximately one range of the discrete scaling factor $e^{\pi/s_0}$~\cite{VonStecherEtAl2008NatPhys}.  The dots are obtained by numerically solving the coupled hyperradial equations, while the solid green curve is obtained from Eq.~(\ref{Kn}) with $N=4$, $\eta=0.01$ and $C_4=55$.  The dashed red curve is a calculation using Eq.~(\ref{Kn}), but ignoring the additional suppression due to $\gamma_{I}$.  
%The three-body parameter $\kappa = \sqrt{2\mu_3|E_{3B}^{(2)}|}/\hbar$, 
%where $E_{3B}^{(2)}$ is the bound-state energy of the second Efimov trimer at unitarity.  
The inset shows the universal phases $\phi$ and $\gamma_{I}$ used in Eq.~(\ref{Kn}). While $\gamma_{I}$ is shown here for $\alpha=10$, results for $K_4^{0^+}$ are independent of $\alpha$ (see text).}
\label{fig:K4Bosons}
\end{center}
%\put(1.0,0.0){\vec{v}}
%\end{picture}
\end{figure}

\emph{Recombination of four bosons with $a<0$.}---Now we specialize our results to the case of four 
identical bosons ($d=9$, $J=0$, $N_{p}=4!$, and no degeneracy.)
Using numerical hyperradial potential curves for four bosons interacting via a short-range model 
potential~\cite{VonStecherEtAl2008NatPhys}, we obtain $K_4^{0^{+}}$ both by solving the coupled channels 
numerically and by using Eq.~(\ref{Kn}) specialized to $N=4$.  In Fig.~\ref{fig:K4Bosons}, we show $K_4^{0^+}$ 
on an absolute scale $[\text{cm}^9/\text{s}]$ (where we use parameters appropriate for ${}^{133}\text{Cs}$: $m=244188$~a.u. and $r_0=100$~a.u.), and in model units of $[\hbar r_0^7/\mu_4]$.
The horziontal axis is shown in units of $r_0$ and in units of the universal ``three-body parameter'' $\kappa=\sqrt{2 \mu_3 |E_{3B}^{(2)}|}/\hbar$, where $E_{3B}^{(2)}$ is the 
bound-state energy of the second Efimov trimer at unitarity.  The overall magnitude of $K_4^{0^+}$ is governed by $-a/r_0$, while the position of the peaks is fixed with respect to $-\kappa a$.  
However, because $\kappa$ and $r_0$ are related by a nonuniversal factor, the relationship between the two horizontal axes in Fig.~\ref{fig:K4Bosons} is model-dependent.
These results show good agreement between the two calculations 
demonstrating the validity of the WKBJ approximation 
%(the dashed red curve is a plot of 
%Eq.(\ref{Kn}) ignoring the additional suppression due to $\gamma_{I}$.)   
The results show the overall $|a|^7$ scaling modified by resonant peaks at:
\begin{equation}
\label{4BosonUniv}
\kappa |a|\approx 0.67 \;\; \text{and} \;\;  \kappa |a| \approx 1.40
\end{equation} 
A cusp in $K_4$ appears at $\kappa |a|\approx 1.56$ when a new atom-trimer channel appears 
(i.e. a new Efimov state becomes bound), whereas peaks appear when a 
four-boson state sits at the threshold of the entrance channel.  

As was recently noted~\cite{VonStecherEtAl2008NatPhys}, 
\emph{two} four-boson states are bound at slightly \emph{less negative} values of $a$ than the values of $a$ at which an Efimov trimer becomes bound.  In the potentials of Fig.~\ref{fig:NBosonSchematic} for $N=4$, the entrance channel is capable of supporting four-boson bound states, but the second trimer-atom channel is not yet available.  
 Because inelastic transitions occur at hyperradii of order the size of the lowest Efimov trimer,  $C$, $\phi$, and 
$\eta$ are universal functions of $a$.  We assume $\eta$ is independent of $a$, and use $\eta=0.01$ and $C_4=55$ in Fig.~\ref{fig:K4Bosons}.  For larger values of $|a|$, the universal structure of Fig.~\ref{fig:K4Bosons} repeats with the three-boson scale factor $e^{\pi/s_0} \approx 22.7$~\cite{VonStecherEtAl2008NatPhys}. The Innsbruck group has recently observed this universal connection between Efimov states and four-boson states and confirmed the spacing implied by Eq.~(\ref{4BosonUniv}), and the observed atom loss 
rates are consistent with the absolute scale in Fig.~\ref{fig:K4Bosons}~\cite{FerlainoEtAl2009PRL}.

In conclusion, we have derived a general formula for the event rate constant for $N$-body recombination.  
The generalized Wigner threshold scaling laws immediately follow, and the overall scaling is resonantly modified by the presence of metastable $N$-body states near threshold.   We then specialize to four-bosons with $a<0$.  

This work was funded in part by the National Science Foundation. 
The authors thank B. D. Esry for extensive discussions related to the present derivations.

%\bibliography{RecombV1}

\end{document}